\documentclass[floatfix,aps,prb,twocolumn,superscriptaddress,showpacs]{revtex4}

\usepackage{graphicx}

\begin{document}
\title{Spin Diffusion in Double-Exchange Manganites}
\author{A. L. Chernyshev} 
\affiliation{Department of Physics and Astronomy, University of
  California, Irvine CA 92697}
\affiliation{Condensed Matter Sciences Division, Oak Ridge National Laboratory,
P.O. Box 2008, Oak Ridge, Tennessee 37831} 
\author{R. S. Fishman} 
\affiliation{Condensed Matter Sciences Division, Oak Ridge National Laboratory,
P.O. Box 2008, Oak Ridge, Tennessee 37831} 

\date{\today}

\begin{abstract}
The theoretical study of spin diffusion in double-exchange magnets by
means of dynamical mean-field 
theory is presented. We demonstrate that the spin-diffusion coefficient
becomes independent of the Hund's coupling $J_H$ in the range
of parameters $J_H S \gg W\gg T$, $W$ being the bandwidth, relevant to
colossal magnetoresistive manganites in the metallic part of
their phase diagram. Our study reveals a close 
correspondence as well as some counterintuitive differences
between the results on Bethe and hypercubic lattices. Our
results are in accord with neutron scattering data and with
previous theoretical work for high temperatures. 
\end{abstract}
\pacs{72.10.-d, 72.15.Lh, 75.30.Vh, 75.40Gb}

\maketitle

Spin diffusion dominates the low-$\omega$, low-${\bf k}$ excitation
spectrum of a magnet in its paramagnetic state and contains important
information about the spin dynamics. In the past, spin diffusion 
was studied intensively in Heisenberg systems and, recently,
has been investigated both theoretically and experimentally for 
strongly-correlated {\it itinerant} magnets.\cite{Kopietz,Bonca,neutrons} 
The current growth of interest in spintronics requires understanding
how local spins relax through their interactions mediated by
itinerant charge carriers rather than through their direct interactions
with each other.\cite{spintronics} 
Among such
systems are the colossal magnetoresistive (CMR) 
manganites which are ferromagnetic metals in a
large part of their phase diagram.\cite{Dagotto} 
Recent systematic neutron scattering experiments on the ferromagnetic CMR
materials revealed a peak centered at $\omega=0$ 
associated with the spin diffusion. This peak was also seen
below the ordering temperature indicating electronic
inhomogeneity with regions having lower $T_c$'s.

In this Letter we present a comprehensive, self-consistent, 
microscopic calculation of spin diffusion applying dynamical mean-field 
theory (DMFT) to the double-exchange (DE)
model.  
We demonstrate that the spin-diffusion coefficient ${\cal D}_s$
is related to the local single-particle Green's
function and can be evaluated as a function of doping and temperature. 
Our results 
agree quantitatively with neutron scattering data on manganites
for a range of doping concentrations.  
Thus, our
approach creates a framework for the self-consistent study of
diffusive spin dynamics in many real materials, including magnetically
doped semiconductors.

Following the general hydrodynamic arguments of Ref.~\onlinecite{hydro}, 
 we write the generalized susceptibility of a
paramagnet for low energies and long wavelengths as
\begin{eqnarray}
\label{chi}
\chi({\bf q},\omega)\simeq \chi({\bf q})\
\frac{{\cal D}_s q^2}{-i\omega+{\cal D}_s q^2}
 \ ,
\end{eqnarray}
which through the fluctuation-dissipation theorem gives the
neutron-scattering dynamical structure factor 
${\cal S}({\bf q},\omega)
\simeq  2 [n_B(\omega)+1]\mbox{Im}\chi({\bf q},\omega)$,
where $\chi({\bf q})$ is the static susceptibility at wave-vector 
${\bf q}$ and  $n_B(\omega)=[e^{\omega/T}-1]^{-1}$. We take
$\hbar=k_B=1$ throughout this paper. 
Further, the generalized susceptibility can be 
related to the spin current-current correlation function using the
dispersion relations\cite{Maleev} and the continuity equation
$\partial S^\alpha({\bf r},t)/\partial t=-\nabla_i \,
j^\alpha_i({\bf r},t)$:
\begin{eqnarray}
\label{chi1}
\chi({\bf q},\omega)=-\frac{q^2a^2}{\omega^2}\biggl[
\Pi
({\bf q},\omega)-\Pi({\bf q},0)\biggr] \ ,
\end{eqnarray}
where $\Pi_{ij}^{\alpha\beta}({\bf q},\omega)=-i\int dt e^{i\omega t}
\theta(t)\langle[j^{\alpha\dag}_i({\bf q},t), j^\beta_j({\bf q},0)] 
\rangle$
is the retarded current-current correlation function\cite{Mahan}, 
$j^\alpha_i({\bf q},t)$ is the $i$th component of the spin
current for the $\alpha$-spin projection, 
$i=1\dots d$, $d$ is the dimensionality,
and $a$ is the lattice constant. We use the isotropy
of the spins above $T_c$ and assume the isotropy of real
space to suppress the indices in
$\Pi_{ij}^{\alpha\beta}({\bf q},\omega)
=\Pi({\bf q},\omega)\delta_{\alpha\beta}\delta_{ij}$.
Combining Eqs. (\ref{chi}) and (\ref{chi1}) in the ${\bf q},\omega
\rightarrow 0$ limit, we write 
the Einstein relation between the spin-diffusion coefficient and the spin
conductivity $\sigma_s$ (which in general is distinct from the
particle conductivity) as 
\begin{eqnarray}
\label{sigma}
{\cal D}_s\chi=\sigma_s=-a^2\lim_{\omega \rightarrow 0}\frac{\mbox{\rm Im} 
[\Pi(0,\omega)]}{\omega} \ ,
\end{eqnarray}
where $\chi=\chi({\bf q}=0)$. These expressions are general and do
not depend on the microscopic model.

We now consider the DE model with Hamiltonian
\begin{eqnarray}
\label{H}
{\cal H}=-t \sum_{\langle ij
\rangle\sigma}\left(c^\dag_{i\sigma}c_{j\sigma}
+\mbox{H.c.}\right)-2J_H\sum_i {\bf S}_{i}\cdot {\bf s}_{i} \ ,
\end{eqnarray}
where $t$ is the nearest-neighbor kinetic energy, $J_H$ is the Hund's
coupling 
between the local Mn$^{3+}$ $S=3/2$ spin and the electronic spin
${\bf s}=c^\dag_{\gamma}
{\mbox{\boldmath$\hat{\sigma}$}}_{\gamma\delta}
c_{\delta}/2$, and 
${\mbox{\boldmath$\hat{\sigma}$}}$ are the Pauli matrices. To describe
the multitude of phases in manganites requires that the orbital,
phonon, or Jahn-Teller terms be included in the above model.\cite{Millis} 
However, the
{\it magnetic} properties of these materials in the metallic part of
their phase diagram, such as the magnetic
excitation spectrum and the ferromagnetic 
transition temperature,\cite{Furukawa,Golosov,Bishop}
are {\it quantitatively} well described by the model in 
Eq. (\ref{H}). Therefore, such a model must 
be also capable of describing spin diffusion in these
systems.\cite{remark2} Regardless of modifications of the
DE model needed to describe a particular real system,
elucidating the dynamic properties of this basic model of
strongly-correlated itinerant magnets
is an important task on its own. 

Within the DE model there is no direct
interaction between the local spins. Thus, the total on-site 
spin ${\bf S}^{tot}_l={\bf S}_l+{\bf s}_{l}$
commutes with the exchange part of Eq. (\ref{H})
and the spin current can be expressed in terms of electronic
operators only:\cite{RF1} $j^\alpha_i({\bf q})=\sum_{\bf k}v^i_{\bf k}
c^\dag_{{\bf k},\gamma}
\hat{\sigma}^\alpha_{\gamma\delta}c_{{\bf k}-{\bf
q},\delta}/2$,
where $v^i_{\bf k}={\nabla^i}
\varepsilon_{\bf k}$ and 
$\varepsilon_{\bf k}=-2t\sum_{i=1}^d \cos k_i a$.
This is simply another way of saying that electrons mediate the
magnetic relaxation processes in an itinerant system.  

The physical
situation relevant to manganites corresponds to strong Hund's
coupling $J_H S \gg W$. Since the
characteristic relaxation time for the electronic spin is short 
and the 
spin relaxation is essentially local, perturbative
approaches to the spin diffusion\cite{Zou} are inapplicable. 
 Therefore, we employ
DMFT, which takes into account the local dynamics in
strongly-correlated systems and has been successfully applied to a
number of problems.\cite{DMFT_review,DMFT_caution}
Using DMFT also simplifies our problem significantly because
the higher-order diagrams in the current-current correlation function of
Eq. (\ref{sigma}), often referred to as vertex corrections, are
identically zero within this approach.\cite{DMFT_review}

Thus, to evaluate the spin-diffusion coefficient we
apply the standard Matsubara formalism to Eq. (\ref{sigma})
using the above definition of the spin current:
\begin{eqnarray}
\label{Ds1}
\frac{{\cal D}_s\chi}{a^2}= \frac{\pi}{2} \sum_{\bf k} (v^i_{\bf k})^2
\int_{-\infty}^{\infty} 
d\nu A_{\bf k}(\nu)^2 \left(-\frac{\partial n(x)}{\partial x}\bigg|_{x=\nu}
\right) \ ,
\end{eqnarray}
where ${\bf k}$ and $\nu$ are the internal momentum and frequency of
the ``bubble'' diagram, respectively. Here $v^i_{\bf k}=2t\sin k_i a$, 
$A_{\bf k}(\nu)=-(1/\pi)\mbox{Im} 
G_{\bf k}(\nu)$ is the electronic spectral function, 
$n(\nu)=[e^{(\nu-\mu)/T}+1]^{-1}$ is the Fermi
function, and $\mu$ is the chemical potential.
 
The DMFT imposes a special form of the Green's function in
which the self-energy is ${\bf k}$-independent and 
is defined from a self-consistency condition
specified below. Within the DMFT parameters of the
model are rescaled such that $\bar{t}=t\sqrt{z}$ is
finite as the dimensionality $d\rightarrow\infty$, where $z=2d$ 
is the number of nearest neighbors.\cite{DMFT_review}  
In the following $\sqrt{2}\bar{t}$ is set to unity.
Generally, the $d=\infty$ limit
is well defined for Bethe and hypercubic lattice geometries.
While the semicircular 
electronic density of states (DOS) of the Bethe lattice is 
convenient for calculations, the Bethe lattice itself lacks the
translational invariance and inversion symmetry
implicitly used to obtain Eq. (\ref{Ds1}). 
Thus, it is important to determine whether
the results on the Bethe lattice are equivalent to the results on
the hypercubic lattice, which is
free from such deficiencies. 
Since many problems have been studied using the Bethe
lattice,\cite{DMFT_review} 
this comparison will have an even broader significance for the DMFT
in general.

We briefly sketch here the DMFT equations for the
DE model.\cite{Furukawa,Auslender} Since the self-energy
is local, one can change $G_{\bf k}(E)\Rightarrow
G_\varepsilon(E)=[E-\varepsilon-\Sigma(E)]^{-1}$, and
$\sum_{\bf k}\Rightarrow\int d\varepsilon\rho_0(\varepsilon)$, where
$\varepsilon=\varepsilon_{\bf k}$, and
$\rho^B_0(\varepsilon)=\sqrt{2-\varepsilon^2}/\pi$ and
$\rho^H_0(\varepsilon)=\exp(-\varepsilon^2)/\sqrt{\pi}$ are the bare DOS's
for Bethe and hypercubic lattices, respectively. 
The properties of the system are obtained from the local Green's function:
\begin{eqnarray}
\label{G_loc1}
g(E)= \int
d\varepsilon\frac{\rho_0(\varepsilon)}{E-\varepsilon-\Sigma(E)} 
\ ,
\end{eqnarray}
where the self-energy is defined from $\Sigma(E)=g_0^{-1}(E)-
g^{-1}(E)$, reminiscent of the Dyson equation, where
$g_0^{-1}(E)$ is the ``Weiss'' function containing the
dynamic influence of the environment on a given local site. The
solution of the single-site problem provides a relation between 
$g_0(E)$ and $\Sigma(E)$.\cite{Furukawa,DMFT_review} 
In the paramagnetic state and in the quasiclassical limit $S\gg 1$, 
such a relation is
particularly simple:\cite{Furukawa} $\Sigma(E)=(J_HS)^2g_0(E)$, which
yields
\begin{eqnarray}
\label{G_loc2}
g(E)= \frac{\Sigma(E)}{(J_HS)^2-\Sigma(E)^2} 
\ .
\end{eqnarray}
 Together with Eq. (\ref{G_loc1}), this gives a
self-consistent condition for $\Sigma(E)$ or $g(E)$. 
\begin{figure}[b]
\includegraphics[angle=270,width=8cm,clip=true]{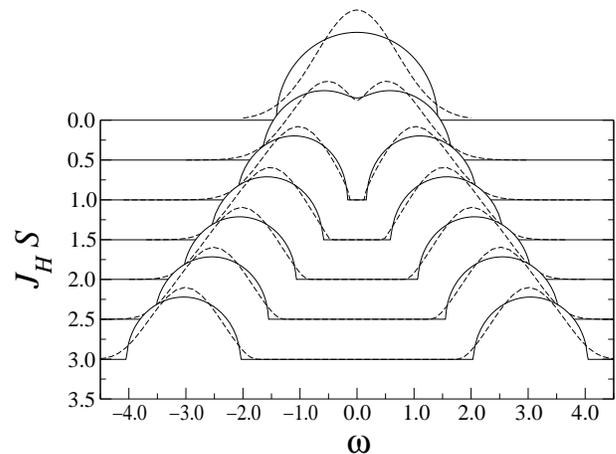}
\caption{Interacting DOS $N(\omega)$ for several values of $J_HS$ for
the Bethe (solid lines) and hypercubic (dashed lines) lattices.} 
\label{fig_1}
\end{figure}

We now compare the Bethe and
hypercubic solution for the single-particle
properties. Fig. \ref{fig_1} shows the evolution
of the interacting DOS $N(\omega)=-(1/\pi)\mbox{Im}g(\omega)$ 
for several values of $J_HS$.
The band splits as $J_H$ increases and 
both lattice geometries exhibit the same qualitative behavior. 
Since
the hypercubic DOS is expected to have 
in-gap states, one may ask whether the
metal-insulator transition is well defined for a half-filled band. 
We find that the band splitting in the Bethe and hypercubic lattices 
happens at the same critical value
$(J_HS)_c=1/\sqrt{2}$. At the transition, the imaginary part of the
self-energy at $\omega=0$ vanishes and the real part diverges. 
As a result,
$N(\omega=0)$ is exactly zero for $J_H>J_H^c$ and the
metal-insulator transition for the half-filled band in the hypercubic
geometry is well defined.\cite{Anokhin} 
At small energies,
$N(\omega)\propto e^{-(J_HS)^4/\omega^2}$ vanishes quite
abruptly.\cite{remark3} In contrast, the ``outer'' tails of the upper and
lower bands behave very similar to the ``bare'' Gaussian form. Note
that for $J_HS\agt 1$, the form of $N(\omega)$ for 
each subband is rather insensitive to the further increase of $J_H$.

We now rewrite Eq. (\ref{Ds1}) for the spin-diffusion coefficient
within the DMFT:
\begin{eqnarray}
\label{Ds3}
\frac{{\cal D}_s\chi}{a^2}= \frac{1}{2z}
\int_{-\infty}^{\infty}
d\nu \left(-\frac{\partial n(x)}{\partial x}\bigg|_{x=\nu} \right) 
\left[\frac{1}{b}-\frac{\partial}{\partial b}\right] \hat{A}(\nu)
\ ,
\end{eqnarray}
where  $b=-\mbox{Im}\Sigma(\nu)$ and 
$\hat{A}(\nu)=\int d\varepsilon\hat{\rho}_0^\alpha (\varepsilon)
A_\varepsilon(\nu)$
with  $\hat{\rho}^{\alpha}_0(\varepsilon)$ being the ``current'' 
DOS defined from the transformation $\sum_{\bf k} \sin^2 k_i a\Rightarrow
\int d\varepsilon \hat{\rho}_0(\varepsilon)$.\cite{Chardopathy} For
the hypercubic lattice 
$\hat{\rho}^H_0(\varepsilon)=\rho^H_0(\varepsilon)/2$ while for the
Bethe lattice
$\hat{\rho}^B_0(\varepsilon)=(2-\varepsilon^2)\rho^B_0(\varepsilon)/3$. 
To obtain Eq. (\ref{Ds3}) we used the relation
$A_{\varepsilon}(\nu)^2=\left[1/b-\partial/\partial b\right]
A_{\varepsilon}(\nu)/2\pi$. 
Since the spin conductivity is proportional to the correlation
function of two spin currents, each scaling as $t\sim 1/\sqrt{z}$, the
prefactor in Eq. (\ref{Ds3}) contains $1/z$.
This means that $d{\cal D}_s\chi$ is finite as $d\rightarrow\infty$,
similar to the particle conductivity.\cite{Metzner}

\begin{figure}[t]
\includegraphics[angle=270,width=8cm,clip=true]{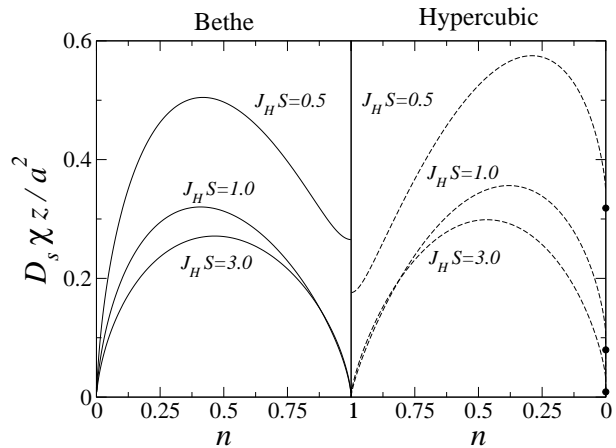}
\caption{${\cal D}_s\chi z/a^2$ versus $n$ 
for several $J_HS$ for the Bethe (solid lines) and hypercubic (dashed
lines) lattices.} 
\label{fig_2}
\end{figure}

In the manganites, the Curie temperature $T_c$ is much smaller than
either the 
Hund's coupling or the bandwidth, in agreement with DMFT
and Monte Carlo calculations for the DE
model.\cite{Furukawa,Furukawa1} 
Therefore, all
realistic temperatures are much smaller than the bandwidth $T\ll W$ and
the derivative of the Fermi-function in the integrand of Eq. (\ref{Ds3})
should be replaced by a $\delta$-function at the chemical
potential. 
Then, combining Eq. (\ref{Ds3})
with the specific form of the DOS's for the Bethe and hypercubic
lattices, one arrives at: 
\begin{eqnarray}
\label{Ds4}
\frac{{\cal D}_s\chi z}{a^2}=
\left\{
\begin{array}{c}
\frac{1}{6\pi}\left(2-\frac{g^{\prime\prime}}{b}
(2-2b^2-f^2)- f g^\prime
\right)^B_{\nu=\mu}  \ , \\
\frac{1}{4\pi} \left(2-\frac{g^{\prime\prime}}{b}
(1-2b^2)-2fg^\prime\right)^H_{\nu=\mu} \ , \ \ \ \ \ 
\end{array}
\right.
\end{eqnarray}
where $g^\prime=\mbox{Re}g(\nu)$,
$g^{\prime\prime}=\mbox{Im}g(\nu)$, and
$f=\nu-\mbox{Re}\Sigma(\nu)$. Thus, 
the spin-diffusion coefficient is expressed through
the local electronic Green's function and self-energy only.
Fig. \ref{fig_2} presents ${\cal D}_s\chi$ as a function of  
the electronic concentration $n$ for 
several $J_HS$. As $n$
varies from 0 to 1, the chemical potential sweeps
from $\omega=-\infty$ to $\omega=0$ in Fig. \ref{fig_1}. 
The results are very similar in both geometries and become
independent of $J_HS$ as $J_HS\rightarrow \infty$ with the maximum
located at $n=0.5$.
In the limit $J_HS\gg~1$, Eq. (\ref{Ds4}) yields a numerical value for
this maximum in ${\cal D}_s \chi z/a^2$: 
$5/6\pi(=0.265)$ and $0.292$ for the Bethe and hypercubic case,
respectively. 
This demonstrates a close quantitative correspondence between
the results in the Bethe and hypercubic lattices, which justifies the
use of the former despite the concerns outlined earlier. An interesting feature 
appears in the results for the hypercubic lattice as $n\rightarrow 0$.
Instead of vanishing, the spin-diffusion coefficient tends to a finite
limit ${\cal D}_s \chi z/a^2
=1/4\pi (J_HS)^2$, shown by circles in Fig. \ref{fig_2}.\cite{Remark4}

In the high-temperature limit $T\gg J_HS\gg 1$ Eq. (\ref{Ds3})
yields the result ${\cal D}_s\chi\sim 1/T$ obtained previously using
a Tchebycheff bounds (TB) formalism.\cite{RF2} 
Numerically, $({\cal D}_s^{DMFT}/{\cal D}_s^{TB})_{Bethe}=
40\sqrt{2}/9\pi^{3/2}=1.13$ 
agree very closely as well. 

It is interesting to analyze our results in the context of CMR
materials. The superexchange (SE) interaction 
is often discussed\cite{Dagotto} as necessary to correct the DE model. 
Within DMFT, the SE coupling $J^{SE}$ must scale as $1/z$. Combining
this with the result of Ref. \onlinecite{Kopietz} one finds that
${\cal D}_s^{SE}\propto 1/z^{3/2}$ is suppressed in comparison with
the DE result ${\cal D}_s\propto 1/z$. 
This also serves as a demonstration that the DE model cannot be simply
reduced to an effective Heisenberg model.
In a real $d=3$
material the role of SE is further reduced by the
smallness of $J^{SE}_{ij}\sim t^2/J_HS$ in comparison with the kinetic
energy $\sim xt$. Therefore, the DE must dominate the 
spin diffusion and one can expect our results to be valid not
only for the metallic part of the CMR phase diagram ($x=0.22..0.5$ for
La$_{1-x}$Ca$_x$O$_3$, $n=1-x$ in Fig.~\ref{fig_2}) but also 
for the ferromagnetic insulating phase ($x\leq 0.22$).\cite{Remark5}
Of course, in the limit $x\rightarrow 0$ DE will
diminish and the SE will dominate. Also, the critical scaling in
the mixed phase $0 < x < 0.12$ (with a mixture of antiferromagnetic
and ferromagnetic orders) can be strongly modified.\cite{Kiselev} 
These cases require separate consideration. At $x\geq 0.5$ 
charge ordering prevents the DE mechanism from being operative. 

\begin{figure}[t]
\includegraphics[angle=0,width=8cm,clip=true]{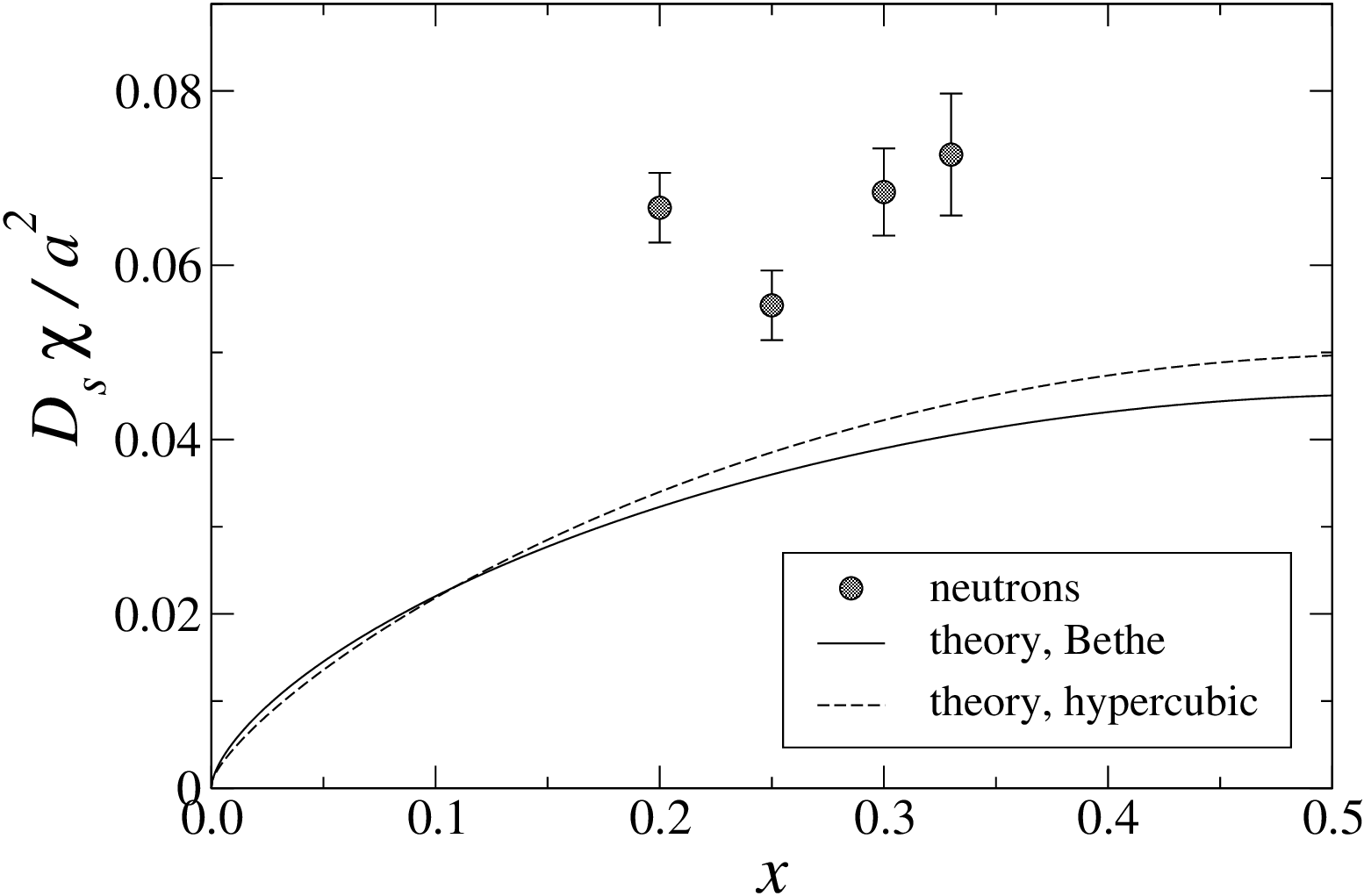}
\caption{${\cal D}_s\chi/a^2$ as a function of $x$.
Experimental values of ${\cal D}_s$ are from Fig. 5 of
Ref.~\onlinecite{neutrons}, $\chi^{exp}$ is described in the text, 
theoretical results are from Eq. (\ref{Ds4}) for 
$J_HS\gg 1$.} 
\label{fig_3}
\end{figure}

A systematic neutron scattering study of the metallic 
manganites recently focused on spin diffusion.\cite{neutrons} 
The width of the observed peak in ${\cal S}({\bf q},\omega)$ centered
at $\omega=0$ scales as $\Lambda q^2$, where 
$\Lambda=2{\cal D}_s$. Experimental results taken at $T^{exp}(x)\simeq
1.1 T_c(x)$, where $T_c(x)$ is the Curie temperature for a given hole
concentration $x$, give $\Lambda=15-30$ meV\AA$^2$. To compare our
results to experiments we use $z=6$ and $a=3.87$ \AA. 
But keep in mind
that spin diffusion persists below the transition point and
that, at least for the experimentally accessible wave-vectors, the
correlation length saturates at about 20 \AA. This 
implies that local magnetic correlations are suppressed by 
electronic inhomogeneities, which are probably associated with the
local charge ordering.\cite{neutrons} 
So for simplicity, we take the susceptibility in the Curie form 
$\chi^{exp}=S^\prime (S^\prime +1)/3T^*$, where $S^\prime(x)=S+(1-x)/2$ is
an average on-site spin and $T^*=T^{exp}(x)$ is known from
Ref.~\onlinecite{neutrons}. Fig. \ref{fig_3}
shows the theoretical data for ${\cal D}_s\chi/a^2$ for $J_HS\gg 1$ from
Eq. (\ref{Ds4}) as a function of $x$  together with
experimental ${\cal D}^{exp}_s\chi^{exp}/a^2$. We note here, that the
theoretical curves in Fig. \ref{fig_3} are virtually independent of
the actual value of $J_HS$ for $J_HS\agt 1$.
This figure 
demonstrates a remarkable agreement between the experimental and 
theoretical results,
which contains no fitting parameters. 
Further improvement of the agreement can be
sought, for example, from taking into account the second $e_g$
band which would effectively reduce $J_H/W$.\cite{remark2}

Since various other experiments also  
indicate the presence of local inhomogeneities in CMR systems,\cite{Adams} 
we propose the following analysis. 
As ${\bf q}$ decreases, long-range magnetic correlations must dominate
$\chi({\bf q})$ and the magnetic correlation length must eventually
exceed the size of the local polaronic distortions.  
So at a fixed temperature close to  $T_c$, $\chi^{exp}({\bf
q})$ will increase and there will be a systematic {\it decrease} 
in the observed value of ${\cal D}^{exp}_s \sim 1/\chi$. 
This set of measurements would provide further information about
magnetic correlations within the inhomogeneities in CMR systems. 

In conclusion, we have presented a self-consistent study of the
spin-diffusion in the double-exchange magnets within the framework of
DMFT. This non-perturbative approach allows us
to calculate the spin-diffusion coefficient at any temperature down to
a transition point. A good agreement with the experiments in the
ferromagnetic CMR manganites and earlier work is
found. Altogether, this provides a new insight into the dynamics of
strongly-correlated itinerant magnets.

We would like to acknowledge valuable discussions with
A.~H.~Castro~Neto, P.~Dai, J.~Fernandez-Baca,  M. Jarrell, M.~N.~Kiselev,
V.~Perebeinos, and A. G. Yashenkin. This research was supported  by 
ORNL,  managed by UT-Battelle, LLC, for the U.S. DOE under
contract DE-AC05-00OR22725.



\begin{thebibliography}{}
\bibitem{remark2} A more realistic model with two
$e_g$ bands becomes effectively equivalent to our single-band
DE model Eq. (\ref{H}) if the
Jahn-Teller splitting or Hubbard repulsion are included.  
Thus, the two bands with splitting
should only change our results quantitatively. 
Our theory can be extended  to the Jahn-Teller
case, but it transforms to a different problem 
for the case of the Hubbard interaction.
\bibitem{DMFT_caution} If the nano-inhomogeneities are truly
important for CMR\cite{Dagotto} the DMFT
will eventually fail to reproduce the
percolative processes for transport in the CMR regime.
\bibitem{Remark4} 
In our problem, diffusion originates from
correlations and the states in the remote tails ($|\mu|\gg J_H S$)
are weakly correlated with diverging relaxation time: 
$\tau\propto e^{\mu^2}$.
The smallness of the DOS cancels this divergence giving a
finite value of ${\cal D}_s\chi\propto \tau N(\mu)$
as $n\rightarrow 0$. 
In this limit the dynamical
spin-diffusion coefficient\cite{RF2} ${\cal
D}(\omega)\chi=\sigma_s(\omega)\propto 1/(\omega^2\tau^2+4)$ has 
negligible width and weight given by $1/\tau$.
\bibitem{Remark5} In fact, the experimental data on ${\cal D}_s$ are very
  similar for the insulating and metallic compounds.\cite{neutrons}
\end{thebibliography}
\end{document}